\documentclass[aps,prd,notitlepage,nofootinbib,preprintnumbers,superscriptaddress]{revtex4}
\usepackage{amssymb}
\usepackage{amsmath}
\usepackage{verbatim}
\usepackage{appendix}
\usepackage[dvipdfmx]{graphicx}
\usepackage[usenames,dvipsnames]{xcolor}
\usepackage{ulem}
\usepackage{tensor}

\begin{document}
\preprint{YITP-16-63, IPMU16-0073}
\title{Is the DBI scalar field as fragile as other $k$-essence fields?}

\author{Shinji Mukohyama}
\affiliation{Center for Gravitational Physics, Yukawa Institute for Theoretical Physics, Kyoto University, 606-8502, Kyoto, Japan}
\affiliation{Kavli Institute for the Physics and Mathematics of the Universe (WPI), The University of Tokyo Institutes for Advanced Study, The University of Tokyo, Kashiwa, Chiba 277-8583, Japan}

\author{Ryo Namba}
\affiliation{Kavli Institute for the Physics and Mathematics of the Universe (WPI), The University of Tokyo Institutes for Advanced Study, The University of Tokyo, Kashiwa, Chiba 277-8583, Japan}

\author{Yota Watanabe}
\affiliation{Kavli Institute for the Physics and Mathematics of the Universe (WPI), The University of Tokyo Institutes for Advanced Study, The University of Tokyo, Kashiwa, Chiba 277-8583, Japan}
\affiliation{Center for Gravitational Physics, Yukawa Institute for Theoretical Physics, Kyoto University, 606-8502, Kyoto, Japan}

\date{\today}

\begin{abstract}
Caustic singularity formations in shift-symmetric $k$-essence and Horndeski theories on a fixed Minkowski spacetime were recently argued. In $n$ dimensions, this singularity is the $(n-2)$-dimensional plane in spacetime at which second derivatives of a field diverge and the field loses single-valued description for its evolution. This does not necessarily imply a pathological behavior of the system but rather invalidates the effective description. The effective theory would thus have to be replaced by another to describe the evolution thereafter. In this paper, adopting the planar-symmetric $1$+$1$-dimensional approach employed in the original analysis, we seek all $k$-essence theories in which generic simple wave solutions are free from such caustic singularities. Contrary to the previous claim, we find that not only the standard canonical scalar but also the DBI scalar are free from caustics, as far as planar-symmetric simple wave solutions are concerned. Addition of shift-symmetric Horndeski terms does not change the conclusion. 
\end{abstract}

\maketitle

\section{Introduction}

The Dirac-Born-Infeld (DBI) action \cite{Born:1934gh,Dirac:1962iy} is a low-energy effective action for a D$p$-brane, the form of which is determined by T-duality and Lorentz invariance of D0-brane. The effective description is valid no matter how large first derivatives of a scalar field corresponding to a position of a brane is, as long as the expression inside the square-root [refer to Eq.~\eqref{eq:LDBI} below for the DBI action] is non-negative and second derivatives thereof are small. If second and/or higher derivatives are also large, corrections of the order of Regge slope $\alpha^\prime$ cannot be negligible, and thus the effective description breaks down.

When describing the motion of a D$p$-brane in a radial direction of extra dimensions, the DBI action without antisymmetric fields is a special case of the action of so-called $k$-essence theory. Such a theory is widely studied in the context of inflation and dark energy \cite{ArmendarizPicon:1999rj,ArmendarizPicon:2000ah}. Recently, $k$-essence theory on a Minkowski background is analyzed by Babichev \cite{Babichev:2016hys} using techniques known in the field of partial differential equations and/or fluid dynamics \cite{Courant:1948}, especially the method of characteristics. Characteristics for a given configuration of a scalar field are hypersurfaces along which perturbations on top of it propagate in the high-frequency and high-momentum limit and correspond to light cones in the case of standard canonical scalar theory. Babichev's analysis assumes the planar symmetry and focuses on the so-called simple wave solutions, in which first derivatives of the scalar field are constant along one of the two families of characteristics. In standard canonical scalar theory, the characteristics are always parallel to each other, and any simple wave solutions (i.e.,~left or right moving modes) travel. In generic $k$-essence theories, however, one can easily construct simple wave solutions in which different characteristics carrying different values of first derivatives of the scalar are not parallel to each other. Then, they intersect with each other within a finite time, and first derivatives of the scalar are no longer single-valued at the intersection. This means that the second and higher derivatives of the field diverge there, and the system exits the regime of validity of the effective description based on the $k$-essence field. The conclusion of the paper \cite{Babichev:2016hys} states that the only caustic-free $k$-essence action in the regime of its analysis may be standard canonical scalar theory.

The present paper further proceeds with the analysis of that work. It is shown that the DBI action is also caustic free as far as the planar-symmetric simple wave solutions in a fixed Minkowski background are concerned. This is consistent with the previous result obtained through the so-called ``completely exceptional'' condition \cite{Lax:1954,Lax:1957,Jeffrey:1964} by Deser, McCarthy, and Sar\i o\~{g}lu \cite{Deser:1998wv}, and therefore, the present paper fills the gap between Refs.~\cite{Babichev:2016hys,Deser:1998wv} and extends the consideration that addition of the higher-order Horndeski terms does not cure the problem of intersecting characteristics. Our result that restricts the form of the action deserves a necessary, but not necessarily sufficient, condition for the effective descriptions of D$p$-branes by the DBI action to be valid, in the sense that second derivatives do not diverge, though the analysis is limited to special solutions, i.e., simple wave solutions.

The rest of the present paper is organized as follows. In Sec.~\ref{sec:review}, we review the analysis of Babichev~\cite{Babichev:2016hys} in a slightly different reasoning. This enables us not only to define notations but also to clarify various steps and concepts such as the coordinate transformation from the standard Cartesian coordinates ($t$, $x$) to the characteristic coordinates ($\sigma_+$, $\sigma_-$), the high-frequency and high-momentum limit of perturbations, categorization of all solutions based on the Riemann invariants, and so on. In Sec.~\ref{sec:linearcharacteristics}, we then seek all shift-symmetric $k$-essence theories in which any planar-symmetric simple wave solutions do not lead to formations of caustics. We find that not only the standard canonical scalar but also the DBI scalar fulfill this criterion. Section \ref{sec:summary} is devoted to a summary of the paper and some discussions. In the Appendix, we show the equivalence between the strategy in Ref.~\cite{Deser:1998wv} and that in the present paper.

\section{Planar caustic singularity}
\label{sec:review}

In this section, we review the analysis of Babichev~\cite{Babichev:2016hys} in a slightly different reasoning, aiming to clarify some physical concepts and procedures. Let us consider a shift-symmetric $k$-essence action for a scalar field $\phi$ in $n$-dimension, 
\begin{equation}
 I=\int \mathrm d^n x \sqrt{-g} \, \mathcal L(X),
\end{equation}
where $X=-\frac{1}{2}g^{\mu\nu}\partial_\mu\phi\partial_\nu\phi$, and we consider nontrivial cases $\mathcal L_X=\mathrm d\mathcal L/\mathrm d X\neq0$. We take the mostly plus signature of the metric, $(-+\dots+)$ and treat the metric as a fixed background. The variation of the action with respect to $\phi$ gives the equation of motion of $\phi$ as
\begin{equation}
 \label{eq:EoM}
 (-\mathcal L_X g^{\mu\nu}+\mathcal L_{XX}\nabla^\mu\phi \nabla^\nu\phi)\nabla_\mu \nabla_\nu \phi=0,
\end{equation}
where $\mathcal L_{XX}=\mathrm d^2 \mathcal L/\mathrm d X^2$ and $\nabla_\mu$ is the covariant derivative.

Let us consider a planar symmetry $\phi=\phi(t,x)$ on the Minkowski metric in the Cartesian coordinates ($t$, $x$, $y$, $z$, $\cdots$). In this case, the equation of motion, Eq.~\eqref{eq:EoM}, is expressed by a system of two equations for two unknown functions $\tau\equiv\dot{\phi}$ and $\chi\equiv\phi^\prime$ of two variables $t$ and $x$,
\begin{align}
 \label{eq:EoM-planar}
 &A\dot{\tau}+2B\tau^\prime+C\chi^\prime=0, \\
 \label{eq:integrability}
 &\tau^\prime-\dot{\chi}=0,
\end{align}
where $\dot{} =\partial/\partial t$, $^\prime=\partial/\partial x$, and
\begin{equation}
 A=\mathcal L_X+\tau^2\mathcal L_{XX},\quad B=-\tau\chi\mathcal L_{XX},\quad C=-\mathcal L_X+\chi^2\mathcal L_{XX}.
\end{equation}
Equation \eqref{eq:integrability} is an integrability condition, i.e.~a necessary and sufficient condition for $\phi$ to exist at least locally with $\tau$ and $\chi$ given. Since $X$ is expressed as $X=\frac{1}{2}(\tau^2-\chi^2)$, $A$, $B$ and $C$ are functions of $\tau$ and $\chi$.

In the following, we consider cases where the equation of motion, Eq.~\eqref{eq:EoM-planar}, is hyperbolic with respect to $\phi$, leading to a single condition 
\begin{equation}
 B^2-AC=\mathcal L_X(2X\mathcal L_{XX}+\mathcal L_X)>0. 
  \label{eq:hyperbilicity}
\end{equation}
This is equivalent to the condition that $\xi_{\pm}$ defined below are real. This is also equivalent to the no-gradient-instability condition $c_s^2>0$ for perturbative modes whose length and time scales are sufficiently shorter than those of the background, where $c_s$ is the speed of sound defined below.

Let us consider a linear combination of Eqs.~\eqref{eq:EoM-planar} and \eqref{eq:integrability},
\begin{equation}
 \label{eq:combination}
 \left[A\partial_t+(2B+\lambda)\partial_x\right]\tau+\left[-\lambda\partial_t+C\partial_x\right]\chi=0,
\end{equation}
where $\lambda$ is an arbitrary function. We now choose $\lambda$ so that the two linear differential operators acted on $\tau$ and $\chi$, respectively, are proportional to each other. Comparing the expressions in the square parentheses in Eq.~\eqref{eq:combination}, we require
\begin{equation}
\frac{2B+\lambda}{A}=\frac{C}{-\lambda},
\end{equation}
which leads to $\lambda=-B\pm\sqrt{B^2-AC}$. The two values of $\lambda$ are different, provided that the hyperbolicity condition (\ref{eq:hyperbilicity}) is satisfied. Then, Eq.~\eqref{eq:combination} becomes
\begin{equation}
 \label{eq:combination2}
 \left(\partial_t+\xi_\pm\partial_x\right)\tau+\xi_\mp\left(\partial_t+\xi_\pm\partial_x\right)\chi=0,
\end{equation}
where
\begin{equation}
 \label{eq:xi}
 \xi_\pm=\frac{B\pm\sqrt{B^2-AC}}{A}
\end{equation}
are functions of $\tau$ and $\chi$. Under the hyperbolicity condition (\ref{eq:hyperbilicity}), the set of equations in \eqref{eq:EoM-planar} and \eqref{eq:integrability} is equivalent to the set of two equations in (\ref{eq:combination2}). Here, for simplicity, we have assumed $A\ne 0$. Relaxing this condition does not change the conclusions of the following analysis.%
\footnote{In the case of $A=0$, one can transform the coordinates, as long as the hyperbolicity \eqref{eq:hyperbilicity} is respected, $B\ne 0$, and follow the same analysis as presented in this section.}

Given the equation of motion, \eqref{eq:combination2}, it is convenient to transform the coordinates ($t$, $x$) into ($\sigma_+$, $\sigma_-$) which satisfy \footnote{In the canonical case $\mathcal L=X$, where $\xi_\pm=\pm1$, $\sigma_\pm$ are null coordinates. The relation between $\partial_{\sigma_\pm}$ and $\partial_t + \xi_\pm \partial_x$ is not unique, and the directions $\partial_{\sigma_\pm}$ are invariant under transforming $\sigma_+$ to a function only of $\sigma_+$ or $\sigma_-$ to a function only of $\sigma_-$, that is, any functions $f(\sigma_+)$ and $g(\sigma_-)$ can serve as the new coordinates, as $\partial / \partial f = f'(\sigma_+) \, \partial_{\sigma_+}$ and $\partial / \partial g = g'(\sigma_-) \, \partial_{\sigma_-}$, thus equally satisfying \eqref{eq:combination2}.}
\begin{equation}
\label{eq:sigma-def}
 \partial_{\sigma_\pm}\propto\partial_t+\xi_\pm\partial_x,
\end{equation}
i.e.,
\begin{equation}
 x_{\sigma_\pm}=\xi_\pm t_{\sigma_\pm},
\end{equation}
where $x_{\sigma_\pm}=\partial x/\partial \sigma_\pm$ and $t_{\sigma_\pm}=\partial t/\partial \sigma_\pm$, provided that the Jacobian $t_{\sigma_+}x_{\sigma_-}-t_{\sigma_-}x_{\sigma_+}=(\xi_--\xi_+)t_{\sigma_+}t_{\sigma_-}$ does not vanish,
\begin{equation}
 \frac{-2\sqrt{B^2-AC}}{A} \, t_{\sigma_+}t_{\sigma_-}\neq0. 
\end{equation}
Curves generated by the vectors $\partial_{\sigma_\pm}$ in the $t$-$x$ plane are called  {\it characteristic curves} or just {\it characteristics}, and we call them $C_\pm$-characteristics.

The physical meaning of $C_{\pm}$-characteristics becomes clear when we consider perturbations in the high-frequency and high-momentum limit, i.e.~in the limit where the time and length scales of perturbations are sufficiently shorter compared with those of the background. In this limit, one can clearly separate the perturbation from the background, and the linearized equation of motion for perturbations $\pi$ reads
\begin{equation}
 A\ddot{\pi}+2B\dot{\pi}^\prime+C\pi^{\prime\prime}\simeq 0,
\end{equation}
where the coefficients $A$, $B$, and $C$ in this equation are their background values. In the high-frequency and high-momentum limit, the geometrical optics approximation is good, and it makes perfect sense to consider a wave of the form $\pi\simeq e^{i(\omega t-kx)}$ with slowly varying amplitude, which leads to\footnote{The Fourier transformation of $\pi$ is not useful due to the $t$- and $x$-dependences of the background.}
\begin{equation}
 A\omega^2-2B\omega k+Ck^2\simeq0.
\end{equation}
Therefore, $\pi$ has dispersion relations
\begin{equation}
 \omega=\xi_\pm k,
\end{equation}
and the velocities of perturbations in the $t$-$x$ plane are $\xi_{\pm}$, meaning that signals propagate along $C_\pm$-characteristics. Here, the phase and group velocities coincide as we have taken the high-frequency and high-momentum limit. All high-frequency and high-momentum signals of $\pi$ follow the curves of $C_\pm$-characteristics at the speed $\xi_\pm$. It is also useful to note that 
\begin{equation}
 \label{eq:addition}
 \xi_\pm=\frac{v\pm c_s}{1\pm vc_s},
\end{equation}
where
\begin{equation}
 \label{eq:v,c_s}
 v=-\frac{\chi}{\tau},\quad c_s=\left(\frac{\mathcal L_X}{2X\mathcal L_{XX}+\mathcal L_X}\right)^{1/2}.
\end{equation}
Equation \eqref{eq:addition} shows that $\xi_{\pm}$ are the relativistic additions of velocities $v$ and $\pm c_s$. The former $v$ represents velocity of fluid elements of $k$-essence $u^x/u^t$, where $u^\mu\propto g^{\mu\nu}\partial_\nu\phi$ (see footnote \ref{footnote:Tmunu}), while the latter $\pm c_s$ represents the speed of sound~\footnote{The energy-momentum tensor of $k$-essence is  $T_{\mu\nu}\equiv\frac{-2}{\sqrt{-g}}\frac{\delta I}{\delta g^{\mu\nu}}=\mathcal Lg_{\mu\nu}+\mathcal L_X\partial_\mu\phi\partial_\nu\phi$. If $\partial_\mu\phi$ is timelike, i.e.~$u_\mu u^\mu=-1$, where $u_\mu\equiv\partial_\mu\phi/\sqrt{2X}$, then $T_{\mu\nu}=\mathcal L(g_{\mu\nu}+u_\mu u_\nu)+(2X\mathcal L_X-\mathcal L)u_\mu u_\nu$, which can be interpreted as a perfect fluid with pressure $P=\mathcal L$ and (rest frame) energy density $\rho=(2X\mathcal L_X-\mathcal L)$. The speed of sound is given by $c_s=(P_X/\rho_X)^{1/2}$.\label{footnote:Tmunu}}, when $\phi^\prime=0$ on the background.\footnote{In general, one cannot take the gauge $t=\phi$ globally maintaining the standard form of the Minkowski metric $\eta_{\mu\nu}$, while this gauge is often convenient in curved spacetime.}

Let us now return to the nonperturbative analysis. We can rewrite the nonperturbative Eq.~\eqref{eq:combination2} by using the characteristic coordinates $\sigma_\pm$ through Eq.~\eqref{eq:sigma-def}, 
\begin{equation}
 \label{eq:combination3}
 \partial_{\sigma_\pm}\tau+\xi_\mp\partial_{\sigma_\pm}\chi=0,
\end{equation}
each of which contains derivatives with respect to only one of the independent variables $\sigma_\pm$. As already stated, the set of these equations is equivalent to the original set of Eqs.~(\ref{eq:EoM-planar}) and (\ref{eq:integrability}), provided that the hyperbolicity condition (\ref{eq:hyperbilicity}) is imposed. Note that $\xi_\pm$ are functions of $\tau$ and $\chi$ and do not depend on their derivatives with respect to $\sigma_\pm$. Each of these equations can be easily integrated along each $C_\pm$-characteristic to give 
\begin{equation}
 \label{eq:invariants}
 \int\frac{\mathrm dX}{Xc_s(X)}\pm\ln\frac{1+v}{1-v}=\Gamma_\pm(\sigma_\mp),
\end{equation}
where $\Gamma_\pm$ are integration constants called {\it Riemann invariants} \cite{Courant:1948}. When deriving Eq. (\ref{eq:invariants}), it is useful to note that 
\begin{equation}
 \mathrm d\left[\int\frac{\mathrm dX}{Xc_s(X)}\pm\ln\frac{1+v}{1-v}\right]
  = \frac{\tau}{Xc_s}
  \left[(1\pm vc_s)\mathrm d\tau + (v\pm c_s)\mathrm d\chi\right],
\end{equation}
and to use the relation (\ref{eq:addition}).

\begin{figure}[t]
\begin{minipage}{0.48\textwidth}
\centering
\includegraphics[width=1\textwidth,clip]{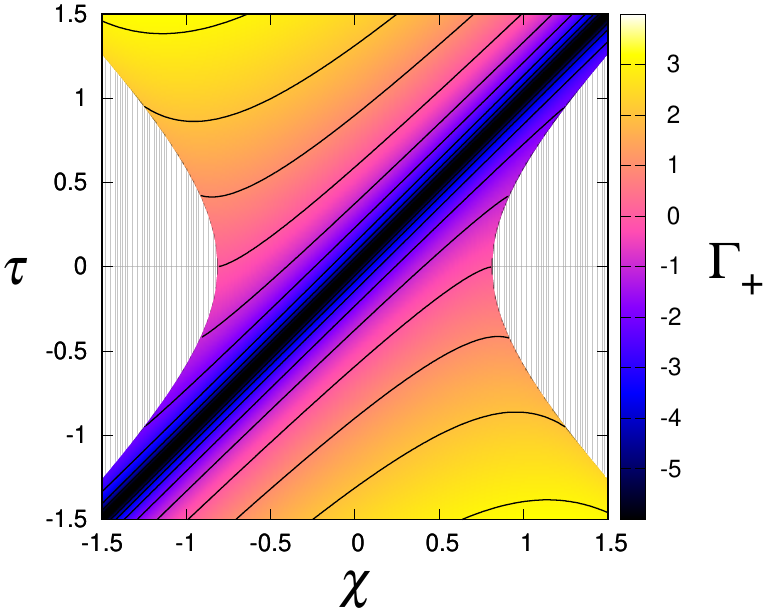}
\caption{Family of $\Gamma_+$-characteristic curves, Eq.~\eqref{eq:invariants}, for $\mathcal L=X+\frac{1}{2}X^2$. The origin of the integration constant $\Gamma_+$ is arbitrary.}
\label{fig:tauchi+(X+0.5X2)}
\end{minipage}
\hfill
\begin{minipage}{0.48\textwidth}
\centering
\includegraphics[width=1\textwidth,clip]{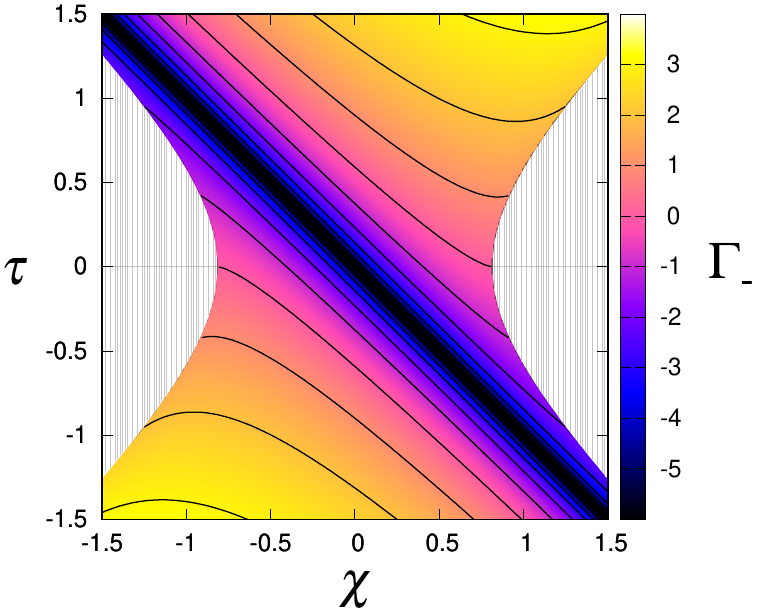}
\caption{Family of $\Gamma_-$-characteristic curves, Eq.~\eqref{eq:invariants}, for $\mathcal L=X+\frac{1}{2}X^2$. The origin of the integration constant $\Gamma_-$ is arbitrary.}
\label{fig:tauchi-(X+0.5X2)}
\end{minipage}
\end{figure}

\begin{figure}[t]
\begin{minipage}{0.48\textwidth}
\centering
\includegraphics[width=1\textwidth,clip]{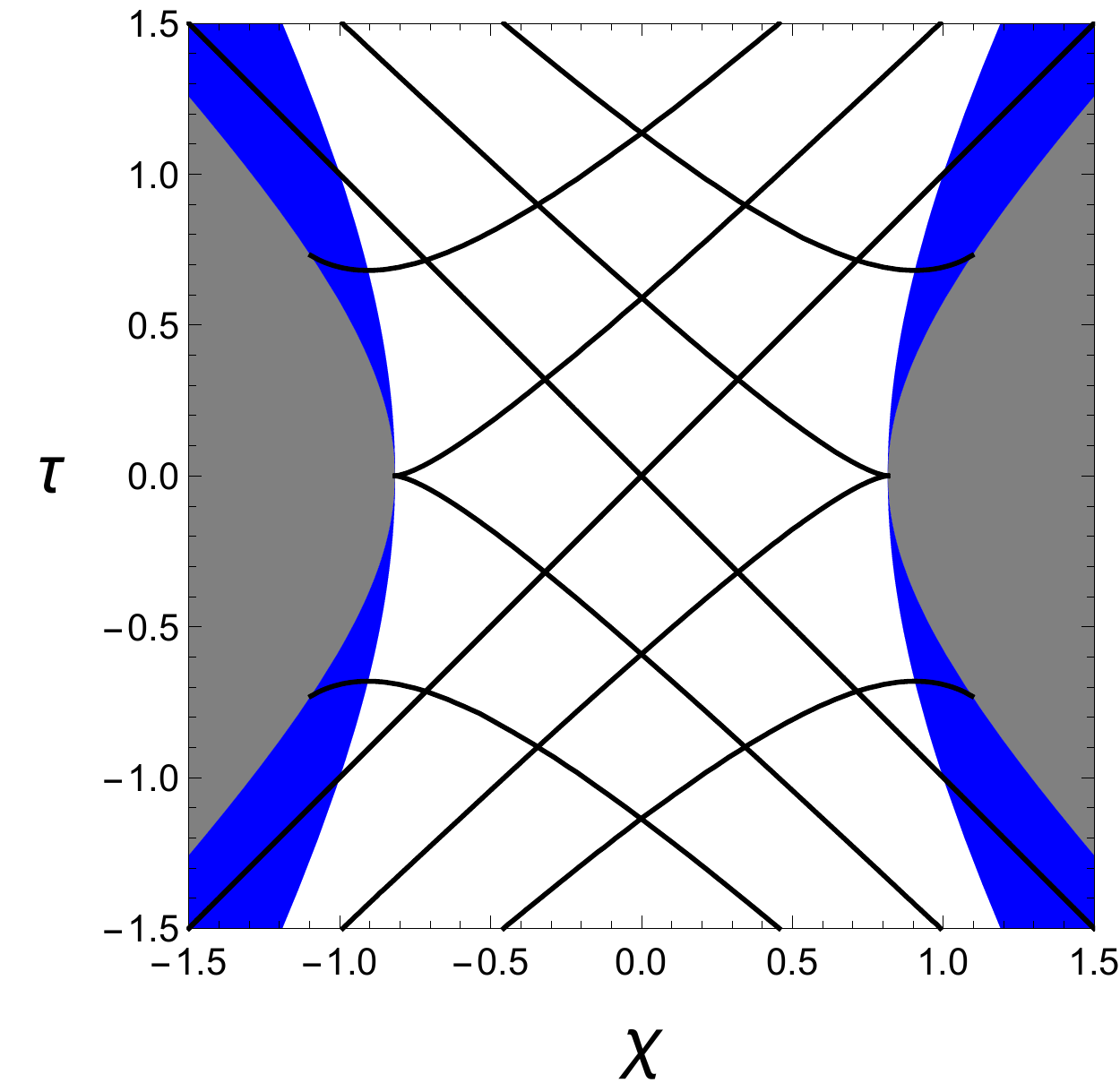}
\caption{Two families of $\Gamma_\pm$-characteristic curves, Eq.~\eqref{eq:invariants}, for $\mathcal L=X+\frac{1}{2}X^2$. In the gray regions, the equation of motion is not hyperbolic and $c_s^2\leq0$. In a solution valued in the blue regions, signs of phase velocities $\xi_\pm$ are the same, and signals propagate only in one direction.}
\label{fig:tauchi(X+0.5X2)}
\end{minipage}
\hfill
\begin{minipage}{0.48\textwidth}
\vspace{4mm}
\centering
\includegraphics[width=0.9\textwidth,clip]{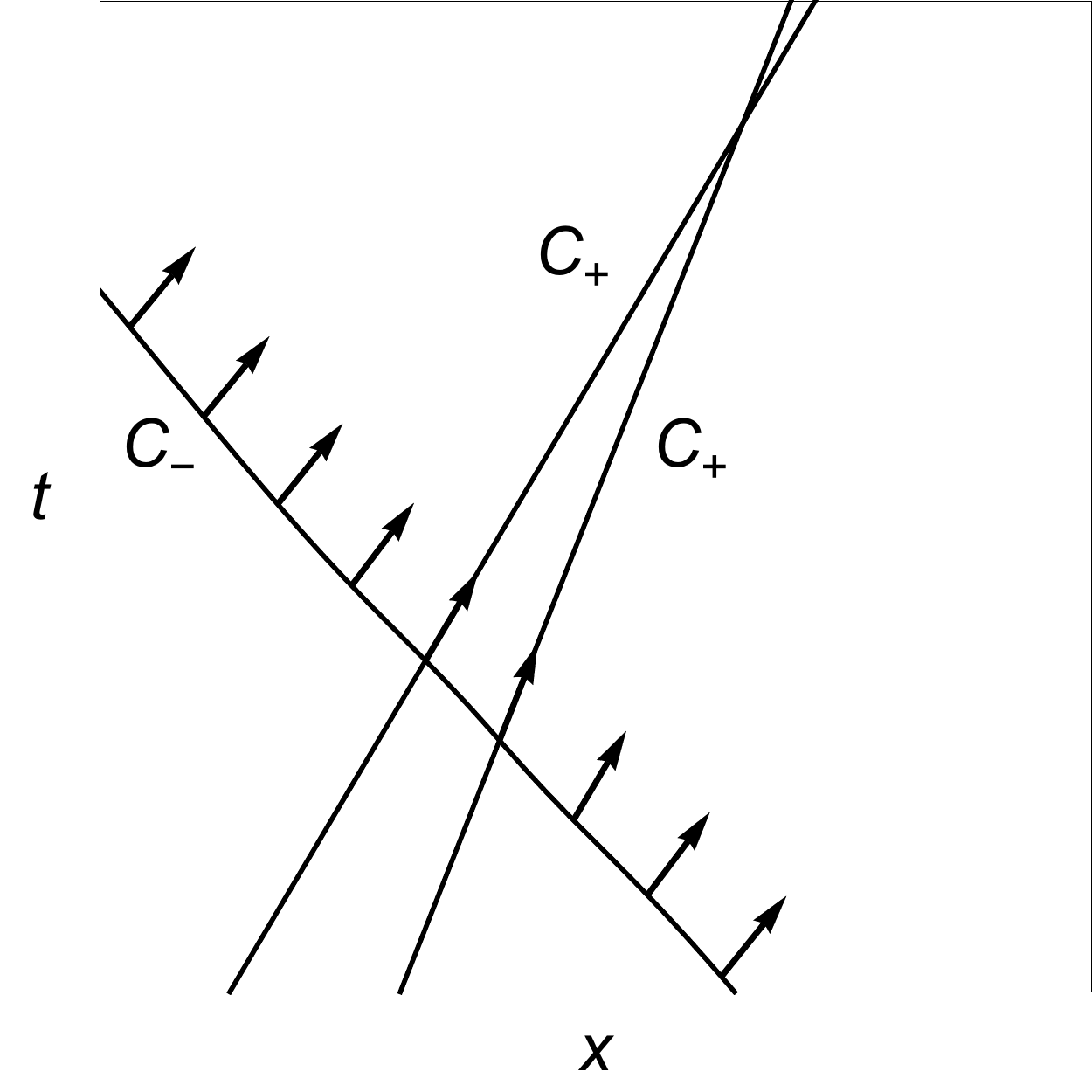}
\caption{{\it Simple wave solution} specified by $\tau>-\chi$ part of $\Gamma_-=\ln2$ in Fig.~\ref{fig:tauchi-(X+0.5X2)} and $\chi=0.7\exp{(-\sigma_-^2)}$ for $\mathcal L=X+\frac{1}{2}X^2$. Two $C_+$-characteristics, along which signals propagate, intersect with each other, and second derivatives of $\phi$ diverge at this point, i.e., caustic, since $\tau$, $\chi$, and thus phase velocities $\xi_\pm$ change along $C_-$-characteristics while they are constant along $C_+$-characteristics.}
\label{fig:caustic}
\end{minipage}
\end{figure}

Since the left-hand side of Eq.~\eqref{eq:invariants} is written entirely in terms of $\tau$ and $\chi$, either equation in (\ref{eq:invariants}) for each fixed value of $\sigma_-$ or $\sigma_+$ defines a curve in the $\tau$-$\chi$ plane. By definition, $\sigma_{\mp}$ [and thus $\Gamma_{\pm}(\sigma_{\mp})$] is constant along each $C_{\pm}$-characteristic for any solutions to the equations of motion (\ref{eq:combination3}). Therefore, Eq.~\eqref{eq:invariants} defines the images of $C_\pm$-characteristics in the $\tau$-$\chi$ plane, which we call $\Gamma_\pm$-characteristics. The images of characteristics in the $\tau$-$\chi$ plane, defined by Eq.~\eqref{eq:invariants}, are independent of solutions in the $t$-$x$ plane.\footnote{Each equation in \eqref{eq:invariants} maps a pair ($\tau$, $\chi$) into one of $\Gamma_\pm$, which is a two-to-one mapping.} See Figs.~\ref{fig:tauchi+(X+0.5X2)}--\ref{fig:caustic}  for the case of $\mathcal L=X+\frac{1}{2}X^2$ as an example. Figures \ref{fig:tauchi+(X+0.5X2)} and \ref{fig:tauchi-(X+0.5X2)} depict the contours in the $\tau$-$\chi$ plane by sweeping the values of $\Gamma_+$ and $\Gamma_-$, respectively. Figure \ref{fig:tauchi(X+0.5X2)} is Figs.~\ref{fig:tauchi+(X+0.5X2)} and \ref{fig:tauchi-(X+0.5X2)} combined, but only for a few values of $\Gamma_\pm$. As can be seen, all the characteristic curves are not straight lines, with the only exception of the ones that pass the point $\tau = \chi = 0$. Figure \ref{fig:caustic} is a solution in the $t$-$x$ plane whose image in the $\tau$-$\chi$ plane is a segment of a $\Gamma_-$-characteristic, showing a few corresponding $C_\pm$-characteristic curves.

Solutions to the equations of motion (\ref{eq:combination3}) fall into three categories \cite{Courant:1948}:
\begin{enumerate}
\item[$(i)$] Firstly, a {\it steady state} is a state where both $\tau$ and $\chi$ are constant. In a region where a solution is steady, $C_\pm$-characteristics are straight lines since $\xi_\pm$ are constant.
\item[$(ii)$] Secondly, a {\it simple wave} is a wave whose image in the $\tau$-$\chi$ plane lies entirely on one of the $\Gamma_\pm$-characteristics. One of the Riemann invariants $\Gamma_\pm$, defined by Eq.~\eqref{eq:invariants}, is constant over the whole $t$-$x$ plane for this type of solution. Suppose that a simple wave is specified by $\Gamma_-=\Gamma_-^0$, where $\Gamma_-^0$ is a constant, and an arbitrary function $\Gamma_+(\sigma_-)$. In this case, the set of two equations in \eqref{eq:invariants} completely determines $\tau$ and $\chi$ as functions of $\sigma_-$ only, meaning that $\tau$, $\chi$, and $\xi_{\pm}$ are all constant along each $C_+$-characteristic. That is to say, an image of a $C_+$-characteristic is a $\Gamma_+$-characteristic, and $\Gamma_-^0$-characteristic intersects with each $\Gamma_+$-characteristic only at a point in the $\tau$-$\chi$ plane, which in turn fixes the values of $\tau$ and $\chi$ on the $C_+$-characteristic. 
\item[$(iii)$] Finally, a {\it general wave} is a wave whose image does not entirely lie on a $\Gamma_\pm$-characteristic, i.e.~neither of the Riemann invariants is constant.
\end{enumerate}
Among them, a simple wave solution can be immediately constructed by specifying a constant value of $\Gamma_-(\Gamma_+)$ and the $\sigma_-$-($\sigma_+$-)dependence of either $\tau$ or $\chi$ [they are independent of $\sigma_+(\sigma_-)$]. Then, both of $\xi_\pm$ are determined as functions of $\sigma_-(\sigma_+)$, and their integrations yield $C_\pm$-characteristics, i.e., paths of signals in the $t$-$x$ plane, given initial positions.

Even among simple waves specified by $\Gamma_-=\Gamma_-^0$, various kinds of solutions can be obtained, depending on how $\xi_+$ changes. Consider, for example, a traveling wave constructed by choosing the $\Gamma_-^0$-characteristic that is linear in the $\tau$-$\chi$ plane. This wave can propagate indefinitely since Eq.~\eqref{eq:combination3}, $\xi_\pm=-\tau_{\sigma_\mp}/\chi_{\sigma_\mp}$, says that $\xi_+$ is the same value for all $C_+$-characteristics. This solution is, however, realized only setting a special condition $\tau=\pm\chi$, i.e.~$X=0$ and $v=\mp1$, in the case of $\mathcal L=X+\frac{1}{2}X^2$ (see Figs.~\ref{fig:tauchi+(X+0.5X2)}--\ref{fig:tauchi(X+0.5X2)}). Note that the lines $\tau=\pm\chi$ are, correspondingly, a $\Gamma_\pm$-characteristic for any $\mathcal L(X)$ as long as ${\mathcal L_X|}_{X=0}>0$ since then $\xi_\mp=\mp1$ from Eq.~\eqref{eq:xi}.%
\footnote{Even if the $\Gamma_\pm$-characteristics $\tau=\pm\chi$ are specified by divergent integration constants $\Gamma_\pm\to-\infty$, they are actually a solution of Eq.~\eqref{eq:combination2} since $(\partial_t+\xi_\pm\partial_x)(\pm\chi)\mp(\partial_t+\xi_\pm\partial_x)\chi=0$.} 
On the other hand, rarefaction and compression waves can also be obtained choosing a curved $\Gamma_-^0$-characteristic, where $\xi_+$ takes different values for different $\sigma_-$ and thus different $C_+$-characteristics. In such a situation, $C_+$-characteristics may intersect with each other within a finite time and form caustics, while $\tau$ and $\chi$ are constant along each $C_+$-characteristic. 
This occurrence is depicted in Fig.~\ref{fig:caustic}, which shows the two $C_+$-characteristic lines intersecting, forming a caustic. 
At caustics, the scalar field cannot be fundamental since its first derivatives are no longer single valued. As a result, the second derivative $\dot{\tau}=\tau_{\sigma_-}(\partial\sigma_-/\partial t)$ diverges since $\partial\sigma_-/\partial t=\xi_+/[(\xi_+-\xi_-)t_{\sigma_-}]$ and $t_{\sigma_-}\to0$ at a caustic, and the same is true for $\tau' = \dot{\chi}$ and $\chi'$. Therefore, second derivatives of $\phi$ diverge, and the system should be described by another theory.\footnote{Such a breakdown of description occurs also when we model a breaking wave in fluid dynamics \cite{Courant:1948}.}

One may speculate that higher-dimensional terms could prevent caustic formations. Such terms often make order of equations of motion higher than two and lead to the Ostrogradsky ghost \cite{Ostrogradsky:1850}, whose Hamiltonian is not bounded from below. Diffeomorphism invariant scalar-tensor theory whose Euler-Lagrange equations are second order in four spacetime dimension is Horndeski theory \cite{Horndeski:1974wa,Deffayet:2011gz,Kobayashi:2011nu}. However, as shown in \cite{Babichev:2016hys}, Horndeski theory on a Minkowski background with above symmetries still admits such a simple wave solution forming caustics. This is because new terms in equations of motion are proportional to $\ddot{\phi} \phi''-(\dot{\phi}^\prime)^2$, and this is zero in a simple wave as
\begin{equation}
 \ddot{\phi}=\dot{\tau}=\frac{\mathrm d\tau}{\mathrm d\sigma_-}\dot{\sigma}_-,\quad
 \phi^{\prime\prime}=\tau^\prime=\frac{\mathrm d\chi}{\mathrm d\sigma_-}\sigma_-^\prime,\quad
 \dot{\phi}^\prime=\frac{\mathrm d\tau}{\mathrm d\sigma_-}\sigma_-^\prime=\frac{\mathrm d\chi}{\mathrm d\sigma_-}\dot{\sigma}_-,
\end{equation}
along $C_+$-characteristics (the same applies for $C_-$-ones).
Terms proportional to $\epsilon^{\mu_1\mu_2\mu\lambda}\tensor{\epsilon}{^{\nu_1}^{\nu_2}^{\nu}_\lambda}\phi_{,\mu_1\nu_1}\phi_{,\mu_2\nu_2}\phi_{,\mu}\phi_{,\nu}$ and $\epsilon^{\mu_1\mu_2\mu_3\mu}\epsilon^{\nu_1\nu_2\nu_3\nu}\phi_{,\mu_1\nu_1}\phi_{,\mu_2\nu_2}\phi_{,\mu_3\nu_3}\phi_{,\mu}\phi_{,\nu}$ in Horndeski theory vanish in the planar-symmetric case $\phi=\phi(t,x)$ because of the antisymmetric nature of $\epsilon^{\mu\nu\rho\sigma}$. Therefore, addition of Horndeski terms does not change the behavior of simple waves in $k$-essence. In other words, a caustic-free subclass of Horndeski theory with the above symmetries should have the $k$-essence part which itself is caustic free.

The effective description by a shift-symmetric $k$-essence or Horndeski field breaks down near caustics. If a theory admits a simple wave solution that forms caustics, then, in order make sense of the theory, one needs to assume the existence of a UV completion (or a partial UV completion) so that a new heavy degree of freedom kicks in before reaching caustics. Babichev suggested such a possibility using a nonlinear sigma model \cite{Babichev:2016hys}.

\section{Theories with linear $\Gamma_\pm$-characteristics}
\label{sec:linearcharacteristics}

In the class of shift-symmetric $k$-essence theories under consideration, let us seek theories in which any planar-symmetric simple wave is a traveling wave and thus does not lead to a formation of a caustic. As the analysis of caustic formation in the previous section demonstrates, a crucial requirement to avoid their formations for simple wave solutions on $\Gamma_-$-($\Gamma_+$-)characteristics in the $\tau$-$\chi$ plane is that all the $C_+$-($C_-$-)characteristic lines are parallel in the $t$-$x$ plane. This implies that the slope $\xi_+$ ($\xi_-$) of the $C_+$-($C_-$-)characteristics must stay constant with changing $\sigma_-$ ($\sigma_+$).\footnote{\label{foot:CE} This is equivalent to the so-called completely exceptional condition that was used in Ref.~\cite{Deser:1998wv} to obtain the same result. The equivalence is proven in the Appendix.} This in turn requires that in the $\tau$-$\chi$ plane, all of $\Gamma_\pm$-characteristics are linear, 
\begin{equation}
\label{eq:linear_tauchi}
 \tau=a\chi+b,
\end{equation}
where coefficients $a$ and $b$ are constants. Lines $\chi=$ const., whose slopes in the $\tau$-$\chi$ plane are infinite, are represented by $a,b\to\infty$ with $b/a$ finite, though such lines correspond to an unphysical infinite phase and group velocities $|\xi_\pm|\to\infty$ due to Eq.~\eqref{eq:combination3}, $\xi_\pm=-\tau_{\sigma_\mp}/\chi_{\sigma_\mp}$ with $\chi_{\sigma_\mp}=0$. Note that the conclusion of this section is the same if $\tau$ and $\chi$ are interchanged. In the following, we demand that for any values of $b$ there exists a value of $a$ such that (\ref{eq:linear_tauchi}) is a simple wave solution.

The differential form of the equation of motion along $\sigma_\pm$, Eq.~\eqref{eq:combination3}, is written as $\mathrm{d} \tau + \xi_\mp \mathrm{d} \chi = 0$ and leads to, with the constraint Eq.~\eqref{eq:linear_tauchi} imposed,
\begin{equation}
 \label{eq:condition}
  (a^2-1)\chi+ab\mp bc_s(X)=0,
\end{equation}
where it is understood that $X$ in this equation is expressed as
\begin{equation}
\label{eq:Xasfunctionofchi}
 X=\frac{1}{2}\left[(a^2-1)\chi^2+2ab\chi+b^2\right] \; .
\end{equation}
We thus demand that for any values of $b$ there exists a value of $a$ such that Eq.~\eqref{eq:condition} is identically satisfied for any $\chi$. Alternatively, since Eq.~\eqref{eq:condition} can be rewritten as 
\begin{equation}
\label{eq:condition2}
b c_s(X) = \pm \sqrt{b^2- 2 \left( 1 - a^2 \right) X} \; ,
\end{equation}
we can demand that for any values of $b$ there exists a value of $a$ such that Eq.~\eqref{eq:condition2} is identically satisfied for any $X$. This restricts the form of $c_s$ as a function of $X$. Once $c_s(X)$ satisfying this demand is obtained, the form of the Lagrangian ${\cal L}(X)$ is related to $c_s$ by using its definition Eq.~\eqref{eq:v,c_s}, as
\begin{equation}
 \label{eq:Lfromc_s}
\frac{\mathrm d}{\mathrm dX}\ln\mathcal L_X=\frac{1}{2X}\left(\frac{1}{c_s^2(X)}-1\right) \; .
\end{equation}
Equation \eqref{eq:Lfromc_s} tells us that once we have $c_s$ as a function of $X$, we can (at least formally) reconstruct ${\cal L}(X)$.

\begin{figure}[t]
\begin{minipage}{0.48\textwidth}
\centering
\includegraphics[width=1\textwidth,clip]{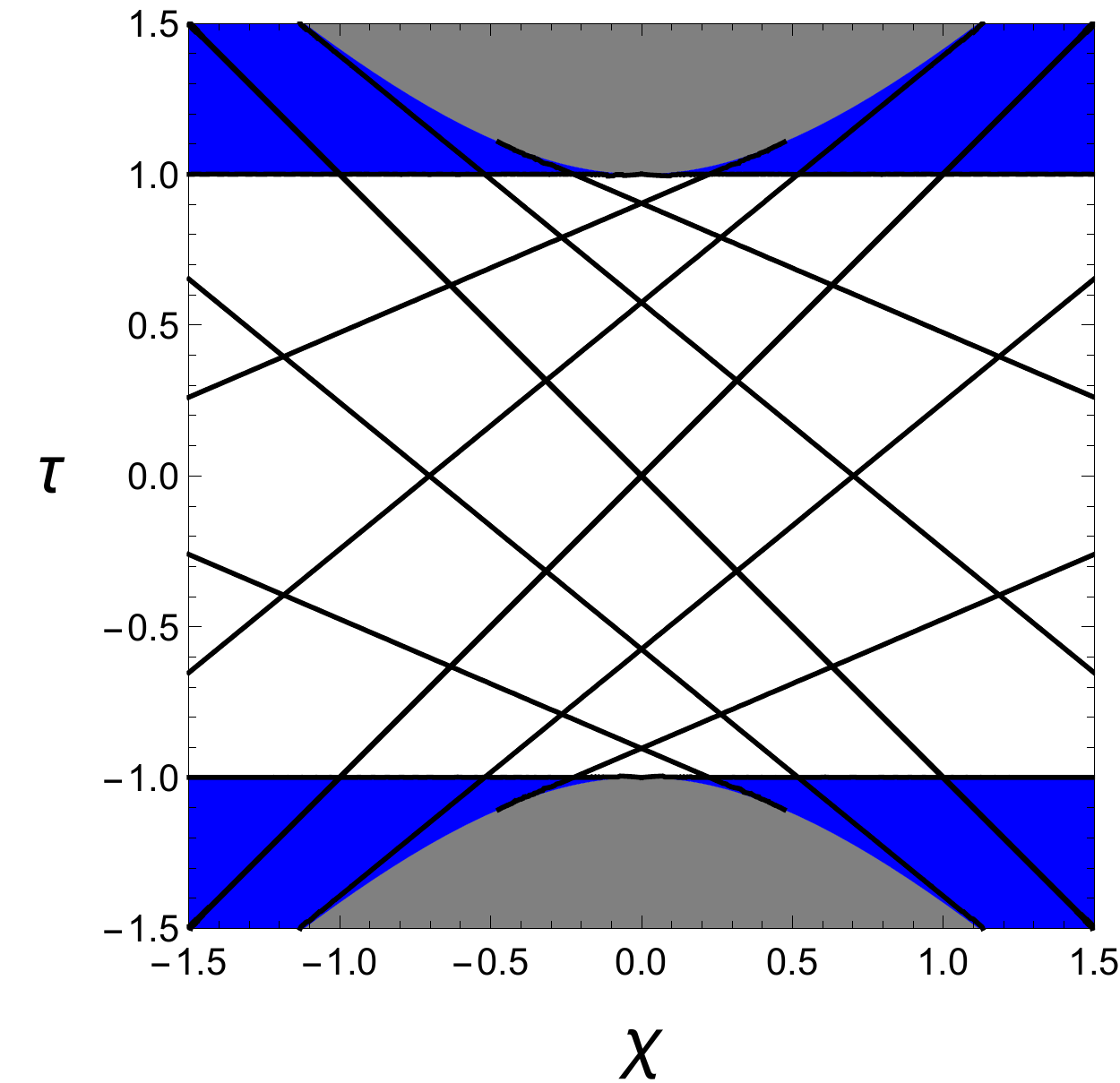}
\caption{Two families of $\Gamma_\pm$-characteristic curves, Eq.~\eqref{eq:invariants}, for $\mathcal L=-\sqrt{1-2X}$. See caption of Fig.~\ref{fig:tauchi+(X+0.5X2)} for gray and blue regions.}
\label{fig:tauchi(-Sqrt[1-2X])}
\end{minipage}
\hfill
\begin{minipage}{0.48\textwidth}
\vspace{4mm}
\centering
\includegraphics[width=0.9\textwidth,clip]{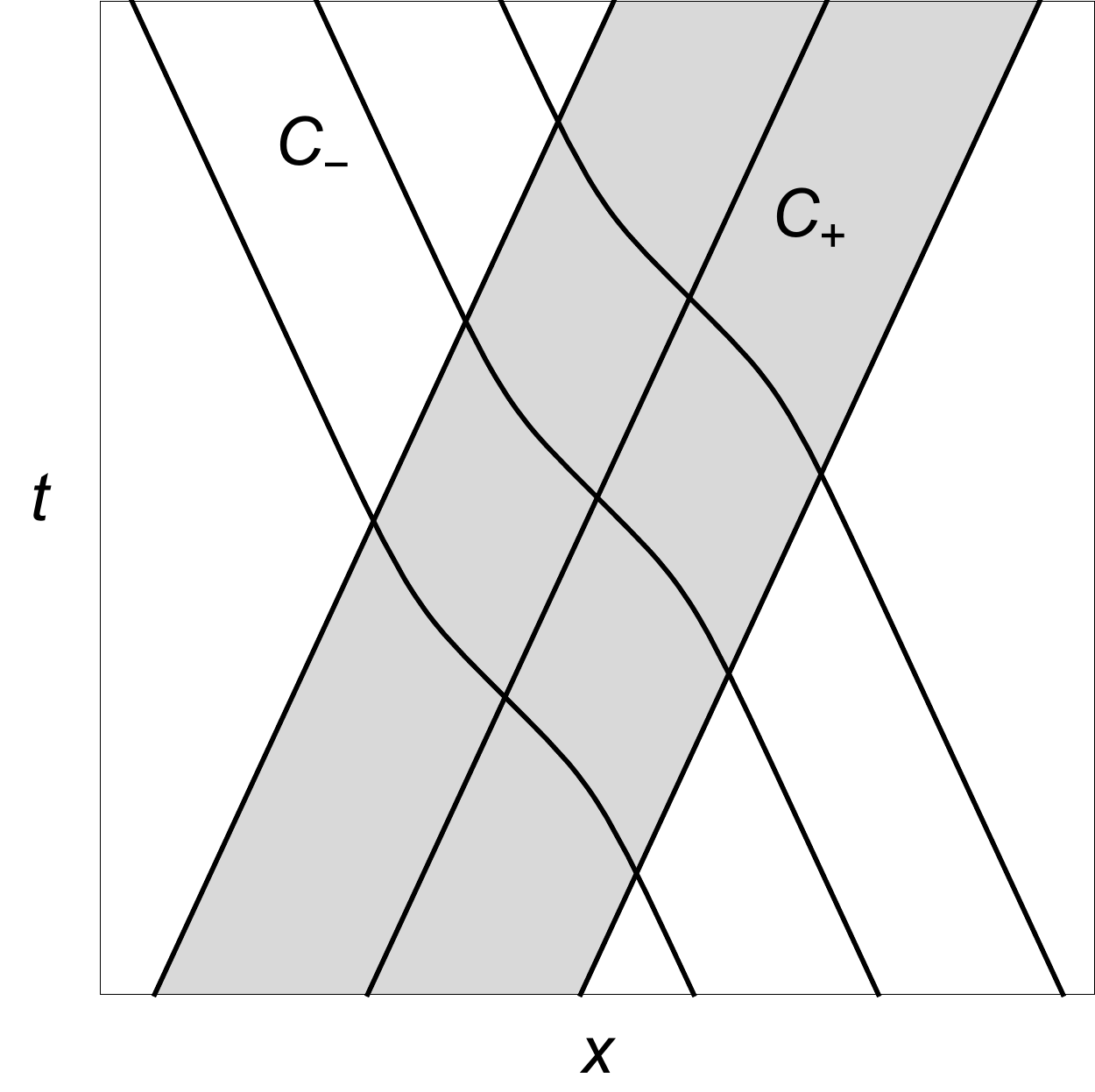}
\caption{{\it Simple wave solution} specified by the $\Gamma_-$-characteristic passing ($\chi$, $\tau$)=(0, $2\sqrt{e}/(1+e)$) and $\chi=0.6\exp{(-\sigma_-^2)}$ for $\mathcal L=-\sqrt{1-2X}$. In the white regions, $\tau$, $\chi$ are almost constant, while they vary in the gray region. Caustics do not form in contrast to Fig.~\ref{fig:caustic}.}
\label{fig:tx(-Sqrt[1-2X])}
\end{minipage}
\end{figure}

Let us now investigate from Eq.~\eqref{eq:Lfromc_s} what forms of ${\cal L}(X)$ are allowed when Eq.~\eqref{eq:condition2} is given, case by case:
\begin{enumerate}
\item[$(i)$] Firstly, in the case where $b=0$, $c_s(X)$ is not constrained, and neither is the form of ${\cal L}(X)$.
\item[$(ii)$] Secondly, in the case where $b\neq0$ and $a^2=1$, we have $c_s^2(X)=1$. This means $\ln\mathcal L_X=$const.; hence, $\mathcal L=e^\text{const.}X+$const, which reduces to (massless) canonical scalar theory $\mathcal L=X$, by a trivial rescaling of $\phi$, plus a cosmological constant term.
\item[$(iii)$] Finally, in the case where $b\ne 0$ and $a^2\ne 1$, Eq.~\eqref{eq:condition2} becomes
\begin{equation}
 \label{eq:c_sDBI}
 c_s^2(X)=1-2fX,
\end{equation}
where $f=(1-a^2)/b^2$. From Eq.~\eqref{eq:Lfromc_s}, we have $\ln\mathcal L_X=\ln\frac{\text{const.}}{\sqrt{1-2fX}}$; hence,
\begin{equation}
 \label{eq:LDBI}
 \mathcal L=-\frac{\text{const.}}{f}\sqrt{1-2fX}+\text{const.},
\end{equation}
which is nothing but the DBI Lagrangian. Indeed, by rescaling the scalar field $\phi$, the Lagrangian (\ref{eq:LDBI}) can be rewritten into the standard form, $\mathcal L=-\lambda\sqrt{1-2X} +\text{const.}$, where $\lambda$ is a constant corresponding to the brane tension and does not contribute to the equations of motion for $\phi$. 
Figure \ref{fig:tauchi(-Sqrt[1-2X])} shows several $\Gamma_\pm$-characteristics in the case of ${\cal L} = - \sqrt{1-2X}$. Actually, they are all straight lines in the $\tau$-$\chi$ plane. Consequently, no characteristics intersect with each other, as can be seen in Fig.~\ref{fig:tx(-Sqrt[1-2X])}, and therefore, no caustics form in any simple wave solution in this theory.
\end{enumerate}

\begin{figure}[t]
\centering
\includegraphics[width=0.5\textwidth,clip]{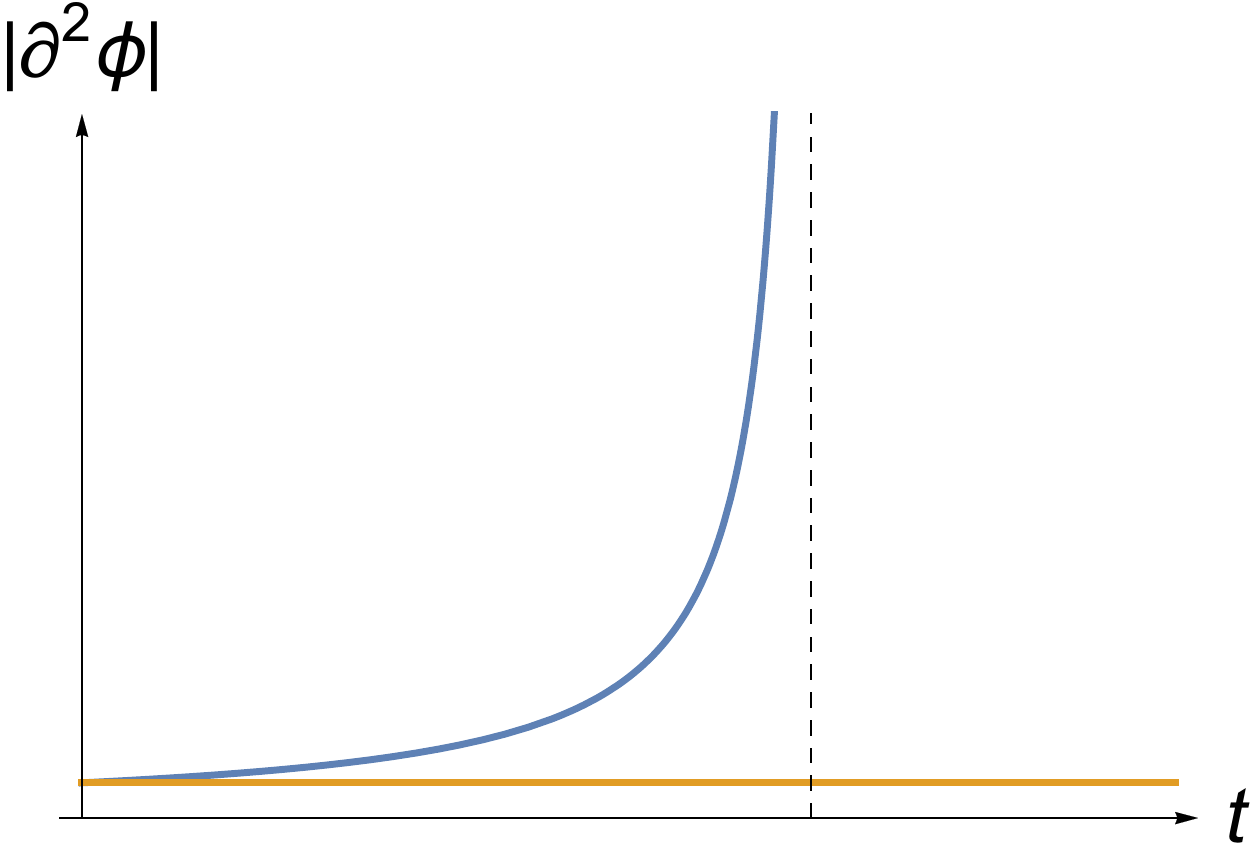}
\caption{Time variation of second derivatives of $\phi$ in a simple wave solution. They diverge in a finite time in a caustic-forming solution, while they are constant in a caustic-free solution.}
\label{fig:d2phi}
\end{figure}

To summarize, we have obtained all shift-symmetric $k$-essence theories in which any planar-symmetric, generic simple wave solutions do not form caustics. Such a class consists of two theories: standard canonical scalar theory and DBI theory. All other shift-symmetric $k$-essence theories admit planar-symmetric simple wave solutions that form caustics. Figure \ref{fig:d2phi} illustrates a schematic comparison of a caustic-forming solution against a caustic-free case. In the former case, the second derivative of $\phi$ diverges at some finite time, from which the theory loses the ability to describe the evolution of the system, while the latter never suffers such a singularity, as far as simple wave solutions are concerned. 

\section{Discussion}
\label{sec:summary}

We have further proceeded the analysis of Ref.~\cite{Babichev:2016hys}, which investigated nonlinear wave solutions with $1$+$1$-dimensional planar symmetry in shift-symmetric $k$-essence theory on a fixed Minkowski background. In the class of solutions called simple wave solutions, signals propagate with constant phase velocities along each characteristic in one of the two families of characteristics, but different characteristics in the family carry different values of the first derivatives (including the phase velocity) of the field. In generic simple wave solutions, different characteristics in the family intersect, and thus, caustics form at the intersection unless the initial condition is fine-tuned on the other family of characteristics. At the intersection, i.e., the caustic, the scalar field fails to be single valued and second derivatives of the field diverge. Thus, the system exits the regime of validity of the effective description based on the $k$-essence field. Thereafter, the effective theory should be replaced by another to describe the evolution of the system. This singularity is not ameliorated by adding higher-dimensional terms of shift-symmetric Horndeski theory, which is free from the Ostrogradsky ghost. The conclusion of Ref.~\cite{Babichev:2016hys} states that the only resolution of the singularity in the regime of its analysis may be to choose Lagrangian to be that of the standard canonical scalar. In the present paper, however, we have found that the class of theories in which generic planar-symmetric simple wave solutions do not form caustics without the fine-tuning includes not only the standard canonical scalar but also the DBI scalar. This is consistent with the previous result obtained through the so-called completely exceptional condition \cite{Lax:1954,Lax:1957,Jeffrey:1964} by Deser, McCarthy, and Sar\i o\~{g}lu \cite{Deser:1998wv} (see the Appendix for the equivalence), and therefore, we have filled the gap between Refs.~\cite{Babichev:2016hys} and \cite{Deser:1998wv}.

It has been known that a fluid of dust, with a vanishing speed of sound $c_s=0$, forms caustics. This is because each fluid element follows a geodesic without communicating with other neighboring elements and different geodesics generically intersect with each other. For this reason, the formation/avoidance of caustics is always one of important issues in any theories with a physical degree of freedom whose speed of sound is zero or small. Examples of such theories include tachyon condensation~\cite{Sen:2002nu,Sen:2002in,Gibbons:2002md}, ghost condensation~\cite{ArkaniHamed:2003uy}, projectable Ho\v rava-Lifshitz theory~\cite{Horava:2009uw,Sotiriou:2009bx}, and so on. In the case of tachyon condensation, the Sen's effective Lagrangian resembles the DBI action that we have found to be caustic free, but the shift symmetry is explicitly broken. The square-root structure of the DBI Lagrangian is multiplied by a field-dependent tension of a decaying brane. As the tachyon condensation proceeds, the tension approaches zero, and the first derivative of the tachyon field becomes large. As a result, the light cone of the effective metric for open-string degrees of freedom becomes narrower and narrower compared with that for closed-string degrees~\cite{Mukohyama:2002vq} and the speed of sound asymptotically vanishes. In realistic cosmological setups this leads to formations of caustics, and thus, the system exits the regime of validity of the effective description in a relatively short time scale~\cite{Felder:2002sv}. In ghost condensation, apparently higher-dimensional but actually leading operators beyond the $k$-essence (and Horndeski) theories inevitably kick in near the (would-be) caustics, and effective fluid elements no longer follow geodesics. Reference \cite{ArkaniHamed:2005gu} found a candidate operator that may avoid formations of caustics and provided some numerical evidences for the caustics avoidance. In the projectable version of Ho\v rava-Lifshitz theory, for the avoidance of caustics, it is necessary to take into account not only the higher spatial derivative terms but also the renormalization group running of the coefficient of a second derivative kinetic term~\cite{Mukohyama:2009tp}. The result of Ref.~\cite{Izumi:2009ry} suggests that such effects should be taken into account also near the central region of a star.

In many (if not all) of those past examples, the codimension-one case %
\footnote{While the occurrence of a planar caustic (or would-be caustic) is at an ($n-2$)-dimensional surface in $n$-dimensional spacetime, the world volume of the corresponding structure is ($n-1$)-dimensional, provided that the system after the caustic formation is evolved by, e.g., the Zel'dovich approximation \cite{Zeldovich:1969sb} or that the would-be caustic is resolved. For this reason, when we discuss evolution or/and resolution of (would-be)	caustics, we consider that a planer caustic has codimension-one.} was the most dangerous regarding formations of caustics, and caustics in higher codimensional cases were relatively easier to resolve. Also, if we include gravitational backreaction in three or higher codimensions, then gravity is strong, and black holes may form before caustics. On the other hand, gravity in the case of codimension-one is much weaker and may be integrable in the sense of Israel's junction condition~\cite{Israel:1966rt}. Hence, we expect that the conclusion in the codimension-one case does not significantly change when we include gravitational backreaction. For these reasons and also for simplicity, in the present paper, we have concentrated on the codimension-one case without taking into account gravitational backreaction. It is nonetheless worthwhile investigating the cases with higher codimensions and the effects of gravitational backreaction in a future work.

In shift-symmetric DBI theory, Eq.~\eqref{eq:c_sDBI}, the speed of sound $c_s$ may become arbitrarily close to $0$, depending on the choice of the initial condition. The vanishing $c_s$ is related to the so-called speed limit, the fact that the non-negativity of the expression inside the square root in the Lagrangian sets the upper bound on the speed of the motion of the D$p$-brane in extra dimensions. As the speed of the brane motion approaches the speed limit, the speed of sound $c_s$ approaches $0$. Despite this fact, the nonlinear analysis in the present paper has shown that caustics never form in shift-symmetric DBI theory, as far as the planar-symmetric simple wave solutions in a fixed Minkowski background is concerned. It is beyond the scope of the present paper to investigate whether canonical and DBI scalar theories are caustic free or not in more general setups.

As already stated several times, only the standard canonical scalar and the DBI scalar fulfill the criterion of caustic avoidance in planar-symmetric simple wave solutions. With this result in mind, it is intriguing to note that standard canonical scalar theory is the leading truncation of the derivative expansion ofDBI theory: $-\sqrt{1-2X}+1=X+\mathcal O(X^2)$. One may thus wonder what would happen if we include the next-to-leading term in the derivative expansion of DBI theory, i.e., $-\sqrt{1-2X}+1=X+X^2/2+\mathcal O(X^3)$. This is exactly the example considered in Sec.~\ref{sec:review}. We have seen that caustics generically form in planar-symmetric simple wave solutions. This conclusion continues to hold up to any finite order in derivative expansion. Nonetheless, if the infinite series in the derivative expansion is resummed to the square-root structure of DBI theory, planar-symmetric simple wave solutions never form caustics. There may be deep physical reasons behind this, but we leave such considerations to a future work.

While the difference between the DBI Lagrangian $-\sqrt{1-2X}$ and the truncated Lagrangian $X+X^2/2$ is small for small $|X|$, planar-symmetric caustics may still form for small $|X|$ in simple wave solutions of the latter theory. This is because each component of the first derivatives of the scalar field can be large in Lorentz signature even if $X$ is close to zero. The difference between two Lagrangians comes at the order of $\mathcal O(X^3)$ at the Lagrangian level and can be small for small $|X|$. However, in the equation of motion shown in Eq.~\eqref{eq:EoM}, $\mathcal L_{XX}$ is multiplied by a product of first derivatives of the scalar field, which may be large. So, the tiny difference in $\mathcal L_{XX}$ of the order of $\mathcal O(X)$ can be enhanced by the product of large first derivatives $\partial^{\mu}\phi\partial^{\nu}\phi$. For this reason, the difference in dynamics may be significant even in the regime with small $|X|$. This can be seen by comparing Figs.~\ref{fig:tauchi(X+0.5X2)} and \ref{fig:tauchi(-Sqrt[1-2X])} for different behaviors of $\Gamma_\pm$-characteristics in the vicinity of $\tau = \pm \chi$, which correspond to small $| X |$ but with large values of $| \tau |$ and $| \chi |$.

As already emphasized, the present paper has dealt with simple wave solutions. In a general wave solution, characteristics in the $t$-$x$ plane are not restricted to be straight lines. There might be cases where characteristics in the $t$-$x$ plane repel before intersecting with each other. There might also be other cases where characteristics in the $t$-$x$ plane tend to attract each other. Whether standard canonical theory and DBI theory are caustic free or not even in such general wave solutions is of future interest but beyond the scope of the present paper.

%

\section*{Acknowledgments}
S.\,M. thanks Alexander Vikman for useful discussions on caustics in modified gravity theories. R.\,N.\, and Y.\,W.\, are grateful to Koji Ichikawa for inspiring discussions at the early stage of this work. All of us thank Gary Gibbons for his bringing the large literature on DBI theory to our attention. This work was supported in part by Japan Society for the Promotion of Science Grants-in-Aid for Scientific Research No.\,24540256 (S.\,M.) and No.\,16J06266 (Y.\,W.). Y.\,W.'s work was supported in part by the Program for Leading Graduate Schools, Ministry of Education, Culture, Sports, Science and Technology (MEXT), Japan. This work was supported in part by World Premier International Research Center Initiative (WPI), MEXT, Japan. 

\appendix
\section{Equivalence to the completely exceptional condition}
 \label{sec:CompletelyExceptional}
In this appendix, we show the equivalence between the so-called ``completely exceptional'' (CE) condition (see Ref.~\cite{Lax:1954,Lax:1957,Jeffrey:1964} for the definition) adopted in Ref.~\cite{Deser:1998wv} and our criterion imposed in Sec.~\ref{sec:linearcharacteristics}. The CE condition, explicitly shown for a scalar field in Ref.~\cite{McCarthy:1999hv}, is written with the notations in the present paper in $1$+$1$ dimensions as
\begin{equation}
 \label{eq:CompletelyExceptional}
 \mathbf{e}_p\cdot \left(
	 \begin{array}{c}
	 \partial_\tau \\
	 \partial_\chi
	 \end{array}
 \right)\lambda_p=0
\end{equation}
for each $p$. Here, $\lambda_p$ and $\mathbf{e}_p$ are eigenvalues and eigenvectors, respectively, of a matrix $\mathbf{M}$ appearing in the equation of motion Eqs.~\eqref{eq:EoM-planar} and \eqref{eq:integrability} when rewritten in the form
\begin{equation}
 \frac{\partial\mathbf{U}}{\partial t}+\mathbf{M}\frac{\partial\mathbf{U}}{\partial x}=\mathbf 0,
\end{equation}
where
\begin{equation}
 \mathbf U= \left(
	 \begin{array}{c}
	 \tau \\
	 \chi
	 \end{array}
 \right), \quad
 \mathbf M=
	 \begin{pmatrix}
	 2B/A & C/A \\
	 -1 & 0
	 \end{pmatrix}.
\end{equation}
Roughly speaking, the eigenvalues $\lambda_p$ correspond to the propagation speed of wavefronts of simple waves, and the CE condition Eq.~\eqref{eq:CompletelyExceptional} states that the gradient of each $\lambda_p$ with respect to the vector $\mathbf U$ is orthogonal to the corresponding propagation direction. This guarantees that no $C_\pm$-characteristics intersect each other, enabling initial wavefronts to evolve without emerging shock waves. Explicitly calculating eigenvalues and eigenvectors of $\mathbf M$, we have
\begin{equation}
 \lambda_\pm=\xi_\pm, \quad \mathbf e_\pm\propto\left(
	 \begin{array}{c}
	 -\xi_\pm \\
	 1
	 \end{array}
 \right),
\end{equation}
where $\xi_\pm$ is defined in Eq.~\eqref{eq:xi}. Then, the CE condition Eq.~\eqref{eq:CompletelyExceptional} is expressed as
\begin{equation}
 \label{eq:CE2}
 \left(\partial_\chi-\xi_\pm\partial_\tau\right)\xi_\pm=0,
\end{equation}
which can be written as
\begin{equation}
 \label{eq:dxi=0}
 \partial_{\sigma_\mp}\xi_\pm=0.
\end{equation}
It is immediate to derive Eq.~\eqref{eq:CE2} from Eq.~\eqref{eq:dxi=0} by regarding $\xi_\pm$ as a function of $\tau$ and $\chi$, using the chain rule and Eq.~\eqref{eq:combination3}, i.e., $\tau_{\sigma_\mp}/\chi_{\sigma_\mp}=-\xi_\pm$. Therefore, the CE condition is equivalent to what was imposed in Sec.~\ref{sec:linearcharacteristics}, as mentioned in footnote \ref{foot:CE}.

\bibliographystyle{apsrmp}

\begin{thebibliography}{99}

 \bibitem{Born:1934gh} 
 M.~Born and L.~Infeld,
 ``Foundations of the new field theory,''
 Proc.\ Roy.\ Soc.\ Lond.\ A {\bf 144}, 425 (1934).
 doi:10.1098/rspa.1934.0059

 \bibitem{Dirac:1962iy} 
 P.~A.~M.~Dirac,
 ``An extensible model of the electron,''
 Proc.\ Roy.\ Soc.\ Lond.\ A {\bf 268}, 57 (1962).
 doi:10.1098/rspa.1962.0124

\bibitem{ArmendarizPicon:1999rj} 
  C.~Armend\'ariz-Pic\'on, T.~Damour and V.~F.~Mukhanov,
  ``$k$-Inflation,''
  Phys.\ Lett.\ B {\bf 458}, 209 (1999)
  doi:10.1016/S0370-2693(99)00603-6
  [hep-th/9904075].

\bibitem{ArmendarizPicon:2000ah} 
  C.~Armend\'ariz-Pic\'on, V.~F.~Mukhanov and P.~J.~Steinhardt,
  ``Essentials of $k$-essence,''
  Phys.\ Rev.\ D {\bf 63}, 103510 (2001)
  doi:10.1103/PhysRevD.63.103510
  [astro-ph/0006373].

\bibitem{Babichev:2016hys} 
 E.~Babichev,
 ``Formation of caustics in k-essence and Horndeski theory,''
 JHEP {\bf 1604}, 129 (2016)
 doi:10.1007/JHEP04(2016)129
 [arXiv:1602.00735 [hep-th]].
 
\bibitem{Courant:1948} 
 R.~Courant and K.~O.~Friedrichs,
 {\it Supersonic Flow and Shock Waves} 
 (Interscience Publ., New York, 1948).

\bibitem{Lax:1954}
 P.~D.~Lax, ``The Initial Value Problem for Nonlinear Hyperbolic Equations in Two Independent Variables,'' Ann.~Math.~Studies {\bf 33}, 211 (1954).
 
\bibitem{Lax:1957}
 P.~D.~Lax, ``Hyperbolic Systems of Conservation Laws II,'' Comm.~Pure Appl.~Math.~{\bf 10}, 537 (1957).

\bibitem{Jeffrey:1964}
 A.~Jeffrey and T.~Taniuti, {\it Non-linear Wave Propagation} (Academic Press, New York, 1964).

\bibitem{Deser:1998wv}
 S.~Deser, J.~McCarthy and \"{O}.~Sar\i o\~{g}lu,
 ```Good propagation' and duality invariance constraints on scalar, gauge vector and gravity actions,''
 Class.\ Quant.\ Grav.\  {\bf 16}, 841 (1999)
 doi:10.1088/0264-9381/16/3/015
 [hep-th/9809153].

\bibitem{Ostrogradsky:1850}
 M.~Ostrogradsky,
 ``M\'emoires sur les \'equations diff\'erentielles, relatives au probl\`eme des isop\'erim\`etres,''
 Mem. Acad. St. Petersbourg {\bf VI 4}, 385 (1850).

\bibitem{Horndeski:1974wa} 
 G.~W.~Horndeski,
 ``Second-order scalar-tensor field equations in a four-dimensional space,''
 Int.\ J.\ Theor.\ Phys.\  {\bf 10}, 363 (1974).

\bibitem{Deffayet:2011gz} 
 C.~Deffayet, X.~Gao, D.~A.~Steer and G.~Zahariade,
 ``From $k$-essence to generalised Galileons,''
 Phys.\ Rev.\ D {\bf 84}, 064039 (2011)
 doi:10.1103/PhysRevD.84.064039
 [arXiv:1103.3260 [hep-th]].

\bibitem{Kobayashi:2011nu} 
 T.~Kobayashi, M.~Yamaguchi and J.~Yokoyama,
 ``Generalized G-inflation: Inflation with the most general second-order field equations,''
 Prog.\ Theor.\ Phys.\  {\bf 126}, 511 (2011)
 doi:10.1143/PTP.126.511
 [arXiv:1105.5723 [hep-th]].

\bibitem{Sen:2002nu} 
  A.~Sen,
  ``Rolling tachyon,''
  JHEP {\bf 0204}, 048 (2002)
  doi:10.1088/1126-6708/2002/04/048
  [hep-th/0203211].

\bibitem{Sen:2002in} 
  A.~Sen,
  ``Tachyon matter,''
  JHEP {\bf 0207}, 065 (2002)
  doi:10.1088/1126-6708/2002/07/065
  [hep-th/0203265].

\bibitem{Gibbons:2002md} 
  G.~W.~Gibbons,
  ``Cosmological evolution of the rolling tachyon,''
  Phys.\ Lett.\ B {\bf 537}, 1 (2002)
  doi:10.1016/S0370-2693(02)01881-6
  [hep-th/0204008].
	
\bibitem{ArkaniHamed:2003uy} 
  N.~Arkani-Hamed, H.~C.~Cheng, M.~A.~Luty and S.~Mukohyama,
  ``Ghost condensation and a consistent infrared modification of gravity,''
  JHEP {\bf 0405}, 074 (2004)
  doi:10.1088/1126-6708/2004/05/074
  [hep-th/0312099].

\bibitem{Horava:2009uw} 
  P.~Ho\v rava,
  ``Quantum Gravity at a Lifshitz Point,''
  Phys.\ Rev.\ D {\bf 79}, 084008 (2009)
  doi:10.1103/PhysRevD.79.084008
  [arXiv:0901.3775 [hep-th]].

\bibitem{Sotiriou:2009bx} 
  T.~P.~Sotiriou, M.~Visser and S.~Weinfurtner,
  ``Quantum gravity without Lorentz invariance,''
  JHEP {\bf 0910}, 033 (2009)
  doi:10.1088/1126-6708/2009/10/033
  [arXiv:0905.2798 [hep-th]].
	
\bibitem{Mukohyama:2002vq} 
  S.~Mukohyama,
  ``Inhomogeneous tachyon decay, light cone structure and D-brane network problem in tachyon cosmology,''
  Phys.\ Rev.\ D {\bf 66}, 123512 (2002)
  doi:10.1103/PhysRevD.66.123512
  [hep-th/0208094].

\bibitem{Felder:2002sv} 
 G.~N.~Felder, L.~Kofman and A.~Starobinsky,
 ``Caustics in tachyon matter and other Born-Infeld scalars,''
 JHEP {\bf 0209}, 026 (2002)
 doi:10.1088/1126-6708/2002/09/026
 [hep-th/0208019].

\bibitem{ArkaniHamed:2005gu} 
 N.~Arkani-Hamed, H.~C.~Cheng, M.~A.~Luty, S.~Mukohyama and T.~Wiseman,
 ``Dynamics of gravity in a Higgs phase,''
 JHEP {\bf 0701}, 036 (2007)
 doi:10.1088/1126-6708/2007/01/036
 [hep-ph/0507120].

\bibitem{Mukohyama:2009tp} 
 S.~Mukohyama,
 ``Caustic avoidance in Ho\v rava-Lifshitz gravity,''
 JCAP {\bf 0909}, 005 (2009)
 doi:10.1088/1475-7516/2009/09/005
 [arXiv:0906.5069 [hep-th]].

\bibitem{Izumi:2009ry} 
  K.~Izumi and S.~Mukohyama,
  ``Stellar center is dynamical in Ho\v rava-Lifshitz gravity,''
  Phys.\ Rev.\ D {\bf 81}, 044008 (2010)
  doi:10.1103/PhysRevD.81.044008
  [arXiv:0911.1814 [hep-th]].

\bibitem{Zeldovich:1969sb} 
 Y.~B.~Zeldovich,
 ``Gravitational instability: An approximate theory for large density perturbations,''
 Astron.\ Astrophys.\  {\bf 5}, 84 (1970).

\bibitem{Israel:1966rt} 
  W.~Israel,
  ``Singular Hypersurfaces and Thin Shells in General Relativity,''
  Nuovo Cim.\ B {\bf 44S10}, 1 (1966)
  [Nuovo Cim.\ B {\bf 44}, 1 (1966)]
  Erratum: [Nuovo Cim.\ B {\bf 48}, 463 (1967)].
  doi:10.1007/BF02710419, 10.1007/BF02712210
  
\bibitem{McCarthy:1999hv} 
 J.~G.~McCarthy and \"{O}.~Sar\i o\~{g}lu,
 ``Shock free wave propagation in gauge theories,''
 Int.\ J.\ Theor.\ Phys.\  {\bf 39}, 159 (2000)
 doi:10.1023/A:1003659520136
 [math-ph/9902004].

\end{thebibliography}

\end{document}